\documentclass[aps,onecolumn,tightenlines,amsmath,amssymb,11pt,superscriptaddress,nofootinbib]{revtex4}

\usepackage{graphicx}
\usepackage{amsmath,amssymb,amsfonts,amsthm,stmaryrd,mathtools,bm,physics}
\usepackage{soul}
\usepackage{xcolor}
\usepackage{tikz}
\allowdisplaybreaks[1]
\usepackage[bookmarks,linktocpage, colorlinks=true, plainpages = false, citecolor = treegreen,  linkcolor=darkblue, urlcolor = darkblue, filecolor = blue]{hyperref} 

\definecolor{tealgreen}{rgb}{0.0, 0.5, 1.0}
\definecolor{darkblue}{rgb}{0., 0.4, 0.8}
\definecolor{cadmiumred}{rgb}{1., 0., 0.22}
\definecolor{treegreen}{rgb}{0., 0.7, 0.3}
\definecolor{emerald}{rgb}{0.31, 0.78, 0.47}
\definecolor{purple}{rgb}{1,0,1}
\definecolor{lime}{HTML}{A6CE39} 

\usepackage{pifont}

\def\be#1\ee{\begin{align}#1\end{align}}

\def\ba{\begin{eqnarray}}
\def\ea{\end{eqnarray}}
\def\nn{\nonumber}

\usepackage[normalem]{ulem}

\begin{document}

\title{Null geodesic defocusing in dynamical black-hole-to-white-hole transitions
}

\author{Johanna Borissova}
\email{j.borissova@imperial.ac.uk}
\affiliation{Abdus Salam Centre for Theoretical Physics, Imperial College London, London SW7 2AZ, United Kingdom}
\author{Stefano Liberati}
\email{liberati@sissa.it}
\affiliation{SISSA, Via Bonomea 265, 34136 Trieste, Italy}
\affiliation{INFN, Sez.~Trieste, Via Valerio 2, 34127 Trieste, Italy}
\affiliation{IFPU--Institute for Fundamental Physics of the Universe, 
Via Beirut 2, 34014 Trieste, Italy}
\author{Matt Visser}
\email{matt.visser@sms.vuw.ac.nz}
\affiliation{School of Mathematics and Statistics, Victoria University of Wellington, PO Box 600, Wellington 6140, New Zealand}

\begin{abstract}
We investigate the defocusing of null geodesics in dynamical, non-singular
black-hole-to-white-hole transitions. Working at the level of spacetime
kinematics, and without assuming any specific gravitational field equations,
we show that the contraction and disappearance of a trapped region, as well as
the subsequent formation and expansion of an anti-trapped region, necessarily
require a violation of the null convergence condition. This conclusion follows
directly from the behaviour of the null expansions across the trapping and
anti-trapping horizons, and is therefore independent of the microscopic
mechanism responsible for singularity resolution. We then illustrate this
general argument by constructing a class of explicit bouncing geometries in
generalised Painlev\'e--Gullstrand coordinates, obtained by promoting static
regular black holes with de Sitter cores to time-dependent black-hole-to-white-hole
transition models. For a Bardeen-type mass function, we show that the required violation of the
null convergence condition is localised within the intermediate
dynamical phase in which the trapped region evaporates and the anti-trapped
region forms. Finally, we argue that the limiting case of an instantaneous
black-hole-to-white-hole transition would require an unbounded violation of the
null convergence condition, signalling a breakdown of the effective continuum
metric description, and the need to appeal to a full quantum-gravitational description.


\end{abstract}

\maketitle
\newpage
\tableofcontents

\bigskip

\section{Introduction}

Gravitational collapse  in general relativity produces singularities under quite generic conditions on the matter~\cite{Penrose:1964wq,Hawking:1970zqf,Hawking:1973uf}, see also~\cite{Senovilla:2014gza}. These conditions are known as energy conditions within the context of the usual Einstein equations~\cite{Barcelo:2002bv, Martin-Moruno:2017exc}. However, the Penrose--Hawking theorems do not actually assume a specific choice of dynamical equations. Instead, they refer to kinematical convergence conditions obtained by contracting the Ricci tensor with a null or timelike vector. 

The kinematic and dynamical mechanisms for avoidance of black hole singularities are an active field of research~\cite{Carballo-Rubio:2025fnc, Buoninfante:2024oxl}. The possible geometric alternatives to producing a singular focusing point during gravitational collapse have been classified in~\cite{Carballo-Rubio:2019nel,Carballo-Rubio:2019fnb}. Excluding the case when  the focusing point is pushed out to infinite affine distance, the kinematical mechanism of singularity avoidance in time-dependent setups, from the point of view of the Penrose theorem, requires null geodesics to defocus, i.e., it requires an intrinsic violation of the null convergence condition (NCC). The latter states that the contraction of the Ricci tensor with any null vector is non-negative. Our main goal here is to establish this statement explicitly for a dynamical bounce describing the evolution from a trapped region into an anti-trapped region through an intermediate untrapped region. Such geometries are well known, for instance, in the context of black-hole-to-white-hole transitions in loop quantum gravity~\cite{Rovelli:2014cta,Haggard:2014rza,Hergott:2022hjm,Hergott:2025elg}.   Even though the gap between the trapped and anti-trapped region may be influenced by quantum fluctuations of spacetime, we will throughout assume that such effects can be parametrised, at least effectively, by means of pseudo-Riemannian geometry --- a key assumption in the geometric classification of regular geometries~\cite{Carballo-Rubio:2019fnb,Carballo-Rubio:2019nel}. In other words, we remain ignorant concerning the particular quantum-gravitational high-energy completion, and consider fluctuations of the geometry to remain small throughout the evolution. We will also comment on when such a semiclassical description can be expected to break down.

Static regular black holes with a de Sitter core, such as the models originally proposed by Bardeen~\cite{Bardeen:1968bh}, Hayward~\cite{Hayward:2005gi},
and Dymnikova~\cite{Dymnikova:1992ux}, do not violate the NCC~\cite{Borissova:2025msp,Borissova:2025hmj}. This is compatible with the Penrose theorem only because such static regular black holes possess an inner/Cauchy horizon, thereby violating the assumption of global hyperbolicity of the theorem. The relevant obstruction to singularity avoidance is then captured instead by the Hawking--Penrose theorem, which involves the timelike convergence condition (TCC), corresponding in general relativity to the strong energy condition (SEC). Indeed, the aforementioned geometries necessarily violate the TCC, as do regular black holes with an anti-de Sitter core~\cite{Borissova:2025hmj}.~\footnote{Note that regular black holes with an anti-de Sitter core also violate the NCC locally near the core, even though they saturate it at the core itself (as does anti-de Sitter spacetime, globally)~\cite{Borissova:2025hmj}.}

By contrast, in dynamical globally hyperbolic spacetimes, the NCC may indeed be violated. A prominent example is provided by black-hole geometries slowly evaporating through Hawking radiation. This should not come as a surprise: intuitively, the contraction of a trapped region requires a local defocusing of null geodesics. As we shall see explicitly, an analogous mechanism is at work in the formation of an anti-trapped region.

\begin{figure}[h!]
	\centering
	\includegraphics[width=0.95\textwidth]{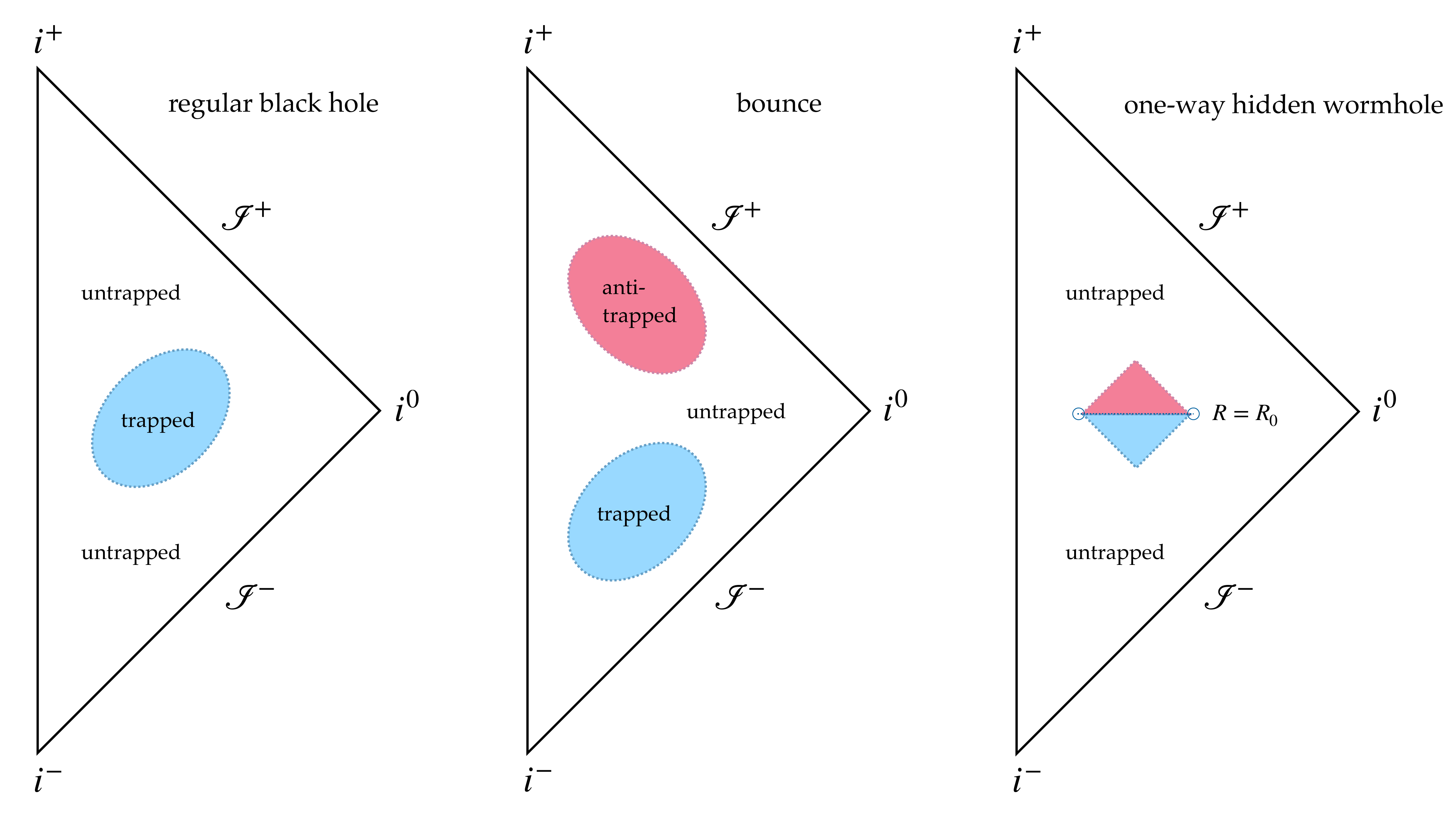}
	\quad \hspace{2cm}
	\caption{\label{Fig:WormholeBounce} Illustration of a dynamical regular black hole, a bounce, and a one-way hidden wormhole.}
\end{figure}

Following this line of reasoning, we will discuss the necessary violation of the NCC during the dynamical formation of a bounce, namely a geometry in which a trapped region evolves into an anti-trapped region through a finite intermediate untrapped region. The presence of this intermediate untrapped region is what distinguishes a bounce from the dynamical occurrence of a one-way hidden wormhole, in which the trapped region transitions instantaneously into an anti-trapped region (see, e.g.,~\cite{Simpson:2018tsi, Lobo:2020ffi, Carballo-Rubio:2019fnb, Carballo-Rubio:2019nel}). While the former class of geometries may admit an effective continuum metric description, the latter may instead signal a breakdown of such a semiclassical description (see, e.g.,~\cite{DAmbrosio:2020mut,Soltani:2021zmv,Ashtekar:2025ptw}), as we will elaborate on later.
Fig.~\ref{Fig:WormholeBounce} illustrates the distinct types of geometries relevant to our discussion.

This article is structured as follows. In Sec.~\ref{Sec:BounceDefinition} we review the geometric notions of a dynamical regular black hole, a one-way hidden wormhole, and a bounce based on the behavior of the expansions of ingoing and outgoing null geodesics. In Sec.~\ref{Sec:PenroseNCC} we review the Penrose theorem and discuss the NCC violation that must be present during the evaporation of a black hole and formation of a  white hole. In Sec.~\ref{Sec:KinematicConstruction} we consider an exemplary kinematical model of a bounce starting from a regular black hole with a de Sitter core to illustrate these general statements.
We finish with a discussion in Sec.~\ref{Sec:Discussion}.

\vspace{2cm}
\section{Dynamical spherically symmetric regular black holes and bounces}\label{Sec:BounceDefinition}

In this section we will review the notions of a trapped, an anti-trapped, and an untrapped region, and therefrom define the three distinct types of geometries relevant for our discussion --- a black hole, a white hole and a wormhole. As we are only interested in the behavior of null geodesics in these spacetimes, we consider a spherically symmetric spacetime in double-null coordinates $(u,v,\theta,\varphi)$ of the form
\ba\label{eq:MetricDoubleNull}
\dd{s}^2 &=& -\dd{u}\dd{v}+ R(u,v)^2 \dd{\Omega}^2\,,
\ea
where $\dd\Omega^2$ is the area element on the unit two-sphere and $R(u,v)>0$ denotes the areal radius function. For further details see the Appendix~\ref{Appendix}.
We refer to $v = \text{const}$  null rays as ingoing, and $u = \text{const}$ null rays as outgoing. These coordinates are particularly useful for describing dynamical black and white holes, and hence also bounces, as well as one-way hidden wormholes. 
To introduce these notions, we consider the two radial null geodesic vector fields
\ba
k_-^\mu \,\,= \,\,\qty( \partial_u)^\mu\,, \quad \quad \quad 
k_+^\mu \,\,=\,\, \qty(\partial_v)^\mu\,.\label{eq:k}
\ea
These geodesics are affine parametrised, i.e., $k_\pm^\nu \nabla_\nu k_\pm^\mu = 0$. Their expansions, $\theta_\pm = \nabla_\mu k^\mu_\pm$, are explicitly given by
\ba
\theta_-\,\, = \,\,  \frac{2}{R} \,\partial_u R\,, \quad \quad \quad 
\theta_+ \,\, = \,\,   \frac{2}{R}\, \partial_v R\,,\label{eq:Theta}
\ea
and they measure the change in transverse area along a constant-$v$ or constant-$u$ slice, respectively. For future reference we also note that the first derivatives of the expansions are given by
\ba
\partial_u \theta_- &=&  - \frac{2}{R^2} \qty(\partial_u R)^2 + \frac{2}{R} \partial_u^2 R \,\, = \,\, -\frac{1}{2} {\theta_-^{2}}  + \frac{2}{R} \partial_u^2 R\,,\label{eq:ThetaDerivative1}\\
\partial_v \theta_+ &=&  - \frac{2}{R^2} \qty(\partial_v R)^2 + \frac{2}{R} \partial_v^2 R  \,\, = \,\, -\frac{1}{2} {\theta_+^{2}}  + \frac{2}{R} \partial_v^2 R \,,\label{eq:ThetaDerivative2}\\
\partial_v \theta_- &=& \partial_u \theta_+  \,\, = \,\,   - \frac{2}{R^2} \qty(\partial_u R)\qty(\partial_v R) + \frac{2}{R} \partial_v \partial_u R \,\, = \,\, - \frac{1}{2}\theta_- \theta_+ + \frac{2}{R} \partial_v \partial_u R \,.\label{eq:ThetaDerivative3}
\ea

A trapped region is characterised by the condition that both ingoing and outgoing
future-directed radial null congruences are converging, while in an
anti-trapped region they are both diverging. In terms of the corresponding
expansions, this translates into
\ba
\text{trapped:} \quad \theta_- < 0\,, \qquad \theta_+ < 0\,,
\\
\text{anti-trapped:} \quad \theta_- > 0\,, \qquad \theta_+ > 0\,.
\ea
Thus, both in trapped and anti-trapped regions, the two expansions have the
same sign, and hence
\ba
\theta_-\theta_+ &>& 0\,.
\ea
By contrast, in the untrapped region surrounding the compact object,
ingoing null rays are converging, whereas outgoing null rays are diverging,
\ba
\text{untrapped:} \quad \theta_- < 0\,, \qquad \theta_+ > 0\,.
\ea
In this case the two expansions have opposite signs, so that
\ba
\theta_-\theta_+ &<& 0\,.
\ea

A dynamical spacetime containing a transient black hole is a geometry in which
a trapped region is surrounded by an untrapped region. Conversely, a dynamical
spacetime containing a transient white hole is a geometry in which an
anti-trapped region is surrounded by an untrapped region. In what follows we
will assume that the black-hole or white-hole core is regular.

A dynamical spacetime containing a one-way hidden wormhole throat, with respect
to the null congruences~\eqref{eq:k}, is characterised by the simultaneous
vanishing of the two expansions at the throat,
\ba
\theta_- = \theta_+ =0\,,
\ea
together with the flare-out conditions
\ba
\partial_u\theta_- \geq 0\,,
\qquad
\partial_v\theta_+ \geq 0\,,
\ea
evaluated at the same surface. In this case the trapped region is joined
directly to an anti-trapped region --- the transition from one to the other occurs
instantaneously at the throat, without an intervening untrapped region.

By contrast, a dynamical spacetime describing a bounce is a geometry in which
a trapped region evolves into an anti-trapped region through a finite
intermediate untrapped region. In terms of the null expansions, the sequence is
\ba
&{}&\qquad\,\,\theta_- < 0\,, \qquad \theta_+ < 0
\qquad \text{trapped}
\nn\\
&{}&\to \quad
\theta_- < 0\,, \qquad \theta_+ = 0
\qquad \text{future inner trapping horizon}
\nn\\
&{}&\to \quad
\theta_- < 0\,, \qquad \theta_+ > 0
\qquad \text{untrapped}
\nn\\
&{}&\to \quad
\theta_- = 0\,, \qquad \theta_+ > 0
\qquad \text{past inner trapping horizon}
\nn\\
&{}&\to \quad
\theta_- > 0\,, \qquad \theta_+ > 0
\qquad \text{anti-trapped}\,.
\ea
Thus, a bounce corresponds to a non-instantaneous black-hole-to-white-hole
transition --- the trapped and anti-trapped regions are separated by a finite
untrapped phase.

\begin{figure}[t]
	\centering
    \includegraphics[width=0.49\textwidth]{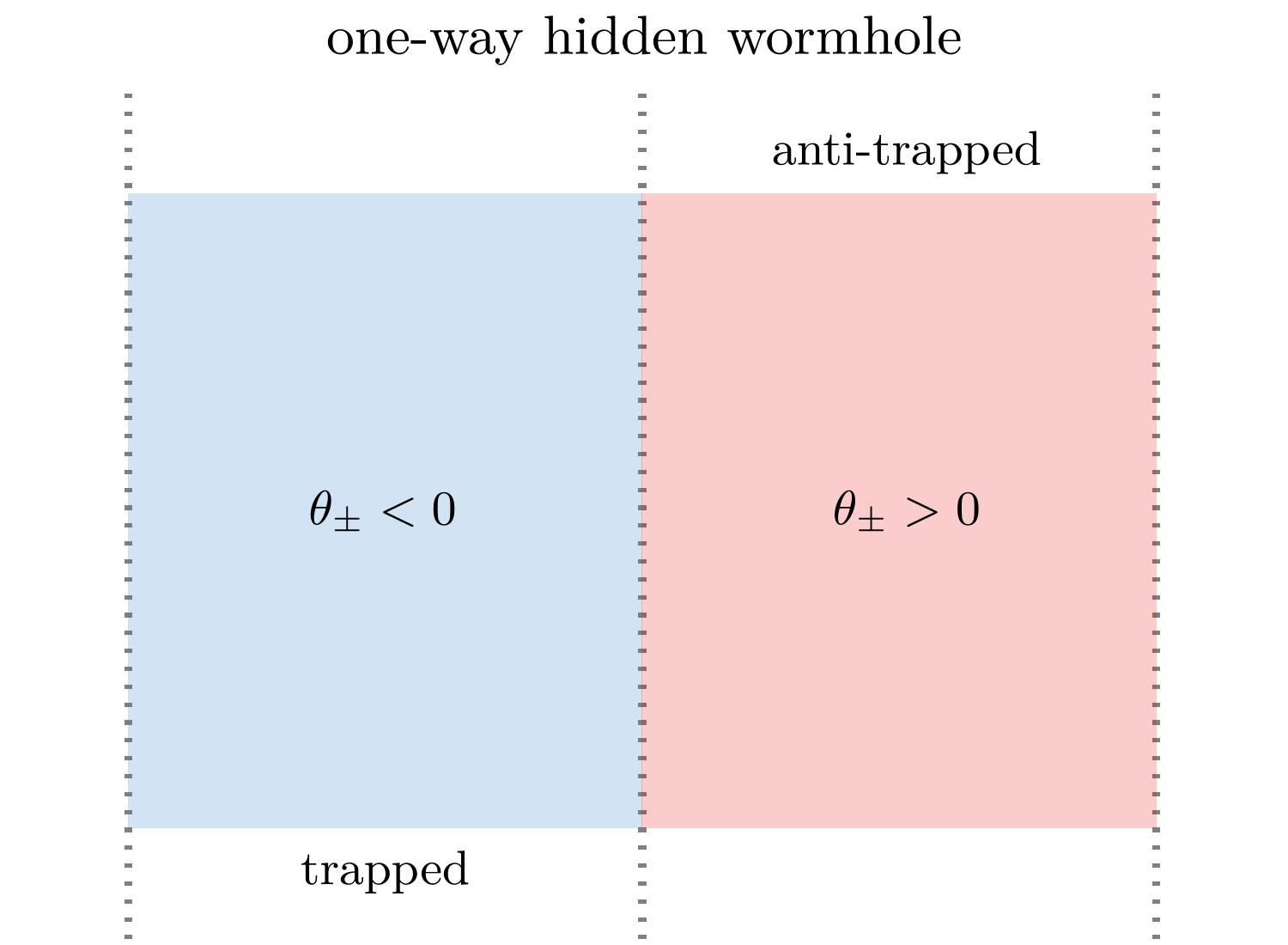}
    \vline
	\includegraphics[width=0.49\textwidth]{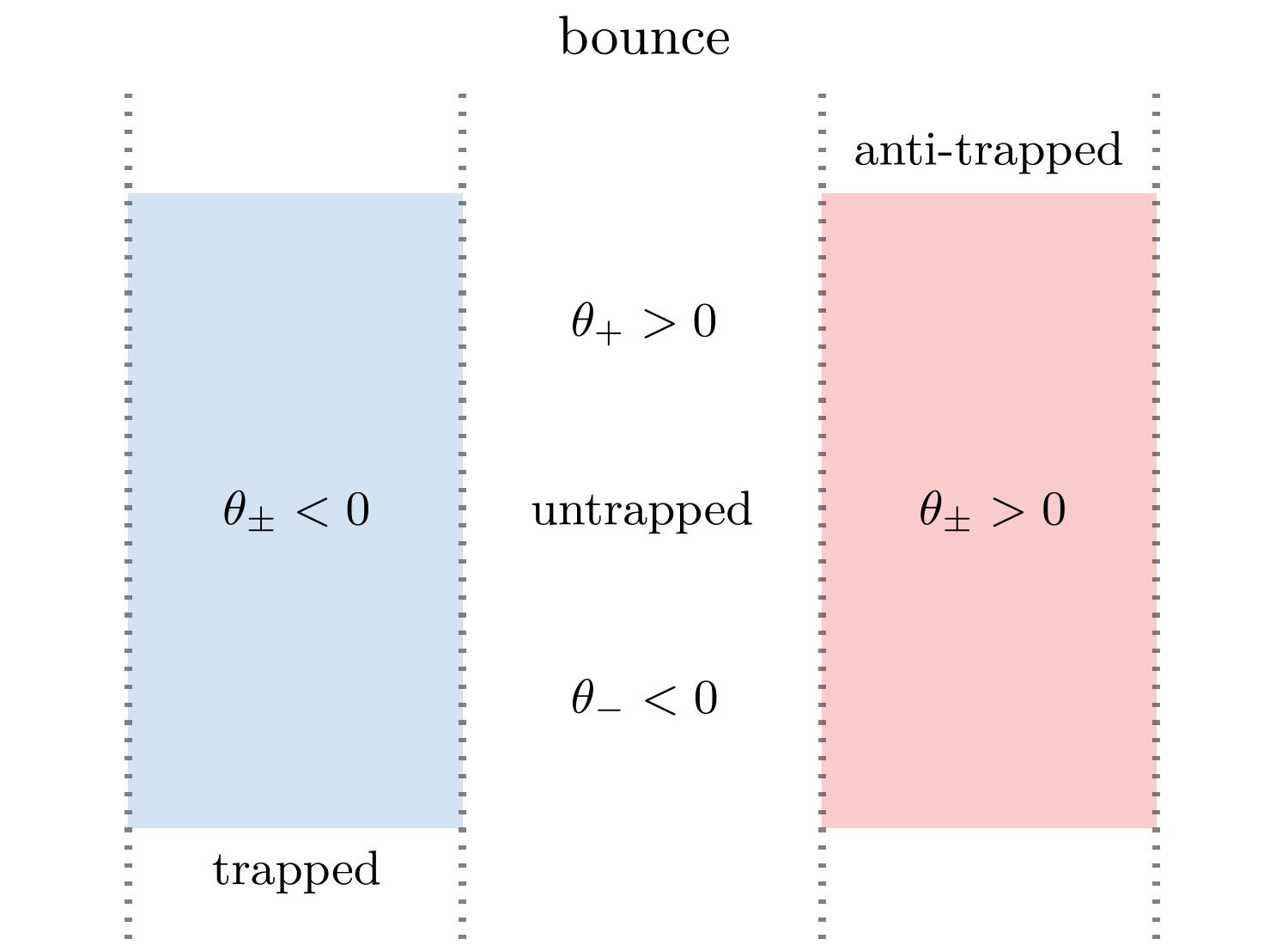}
	\caption{\label{Fig:BounceChronology} Chronology of a one-way hidden wormhole versus that of a bounce. Time is increasing from left to right in both figures. The dotted lines denote the outer/inner horizons bounding the trapped/anti-trapped regions. The outer and inner horizon of the trapped region are more precisely characterised, respectively, as future outer/inner trapping horizons, while those of the anti-trapped region are, respectively, past outer/inner trapping horizons.}
\end{figure}

Fig.~\ref{Fig:WormholeBounce} shows an illustration of a dynamical regular black hole, a bounce and a one-way hidden wormhole. Fig.~\ref{Fig:BounceChronology} summarises pictorially the chronological behaviour of the expansions $\theta_\pm$ for a one-way hidden wormhole versus a bounce.

It should also be noted that the in the process of formation or evaporation of a trapped or anti-trapped region, the spacetime geometry passes through an extremal state in which the outer and inner horizons coincide, i.e., at this instant of time there exists only a single extremal horizon.~\footnote{Our assumption of a transient trapped region implies that the dynamics underlying the evaporation of a regular black hole cannot be described by standard Hawking evaporation, according to which the extremal endpoint of an evaporating regular black hole cannot be reached in finite time. This ultimately implies that the key assumption of adiabadicity in the derivation of Hawking radiation must be violated, cf.~\cite{Arrechea:2026rua} for related recent discussions.}

\section{Penrose theorem and null convergence condition}\label{Sec:PenroseNCC}

To proceed, we  first review the central input assumptions of the Penrose 1965 singularity theorem~\cite{Penrose:1964wq,Hawking:1973uf}. This theorem establishes the null geodesic incompleteness of a spacetime $(\mathcal{M},g)$ under the following conditions,
\begin{itemize}
	\item[i)]{null convergence condition (NCC): $R_{\mu\nu}k^\mu k^\nu \geq 0  \,\,\,\forall\, k^\mu $ null vectors,}
	\item[ii)]{existence of a non-compact Cauchy hypersurface $\Sigma$ in $\mathcal{M}$,}
	\item[iii)]{existence of a closed trapped surface $\mathcal{T}$ in $\mathcal{M}$.}
\end{itemize}

Evaluating the NCC for the two null vectors $k^\mu_\pm$ in~\eqref{eq:k} results in
\ba
R_{\mu\nu} k_-^\mu k_-^\nu &=& - \frac{2}{R} \partial_u^2 R \,\,\geq \,\, 0\,,\label{eq:RkMinus}\\
R_{\mu\nu} k_+^\mu k_+^\nu &=& - \frac{2}{R} \partial_v^2 R \,\, \geq \,\, 0\label{eq:RkPlus}\,.
\ea
The definition of a wormhole throat in Sec.~\ref{Sec:BounceDefinition} as a surface where $\theta_\pm = 0$, and moreover $\partial_u \theta_- \geq 0$ and $\partial_v \theta_+ \geq 0$, assuming strict inequality in the latter flare-out conditions, implies that a wormhole throat represents a strict non-zero local minimum of the areal radius function $R$. The NCC is thus manifestly violated at a wormhole throat. See also the related discussion in~\cite{Borissova:2025msp}. For future reference, we also state explicitly the cross-contraction of the Ricci tensor with the two null vectors in~\eqref{eq:k},~\footnote{This quantity can be written as
\ba
R_{\mu\nu} k_-^\mu k_+^\nu &=& \frac{1}{2} \qty[R_{\mu\nu} (k_-^\mu+k_+^\mu)(k_-^\nu+ k_+^\nu) - R_{\mu\nu} k_-^\mu k_-^\nu - R_{\mu\nu} k_+^\mu k_+^\nu ] \\
&=& \frac{1}{2} \qty[R_{\mu\nu}v^\mu v^\nu - R_{\mu\nu} k_-^\mu k_-^\nu - R_{\mu\nu} k_+^\mu k_+^\nu ]\,,
\ea
where $v^\mu = k_-^\mu +k_+^\mu$ is a timelike vector. Thus, it measures how far the TCC ($R_{\mu\nu}v^\mu v^\nu \geq 0$ for $v^\mu$ timelike) deviates from the antipodally averaged NCC.}
\ba
\label{eq:RkMinusPlus}
R_{\mu\nu} k_-^\mu k_+^\nu &=& -\frac{2}{R} \partial_u \partial_v R\,.
\ea

To investigate the NCC for a transient dynamical black hole, and more generally for a bounce where the black hole is followed by a white hole, we will first focus on the behavior of the expansion $\theta_+$ from the interior of the trapped region up to the inner trapping horizon which defines the future boundary between the trapped and untrapped region. Across this boundary, $\theta_+$ changes sign from negative to positive, in particular along the $v$-direction.  

For a bounce describing the transition of the black hole into a white hole, we will in addition focus on the behavior of the expansion $\theta_-$ at the inner anti-trapping horizon which defines the past boundary between the untrapped and anti-trapped region. Across this boundary, $\theta_-$ changes sign from negative to positive, in particular along the $u$-direction. 

From the expressions for the expansions in~\eqref{eq:Theta}, their derivatives in~\eqref{eq:ThetaDerivative1}--\eqref{eq:ThetaDerivative2}, and the expressions of the Ricci tensor contracted with each of the null vectors~\eqref{eq:RkMinus}--\eqref{eq:RkPlus}, it follows that
\ba
\partial_u \theta_-& = & - \frac{1}{2}{ \theta_-^{2}}  - R_{\mu\nu} k_-^\mu k_-^\nu \,,\label{eq:RaychaudhuriThetaMinus}\\
\partial_v \theta_+& = & - \frac{1}{2}{ \theta_{+}^{2}} - R_{\mu\nu} k_+^\mu k_+^\nu \,.\label{eq:RaychaudhuriThetaPlus}
\ea
 When the trapped region of a dynamical black hole shrinks, i.e., the inner horizon expands and the outer horizon contracts, at the boundary of the trapped region where $\theta_+=0$, for any slice of constant $u=u_*$ that intersects the trapped region, it holds
\ba
\partial_v \theta_+(u_*,v) &>& 0\,\,\,\quad \quad \Longleftrightarrow \,\,\, \quad \quad R_{\mu\nu}k_+^\mu k_+^\nu(u_*,v) \,\, < \,\, 0\,.
\ea
That is, the contraction of a trapped region requires an intrinsic violation of the NCC. 

Similarly, when an anti-trapped region is formed from an untrapped region, i.e., the inner horizon moves inwards and the outer horizon expands, at the boundary of the anti-trapped region where $\theta_- =0$, for any slice of constant $v=v_*$ that intersects the anti-trapped region, it  holds
\ba
\partial_u \theta^-(u,v_*) &>& 0\,\,\,\quad \quad \Longleftrightarrow \,\,\, \quad \quad R_{\mu\nu}k_-^\mu k_-^\nu(u,v_*) \,\, < \,\, 0\,.
\ea
That is, the formation of an anti-trapped region requires an intrinsic violation of the NCC. Thus, independently, the evaporation of a trapped region, the formation of an anti-trapped region, and hence also a dynamical bounce, necessarily violate the NCC.

\subsection{Contraction and evaporation of a trapped region}\label{SecSub:Trapped}

\begin{figure}[h!]
	\centering
	\includegraphics[width=0.95\textwidth]{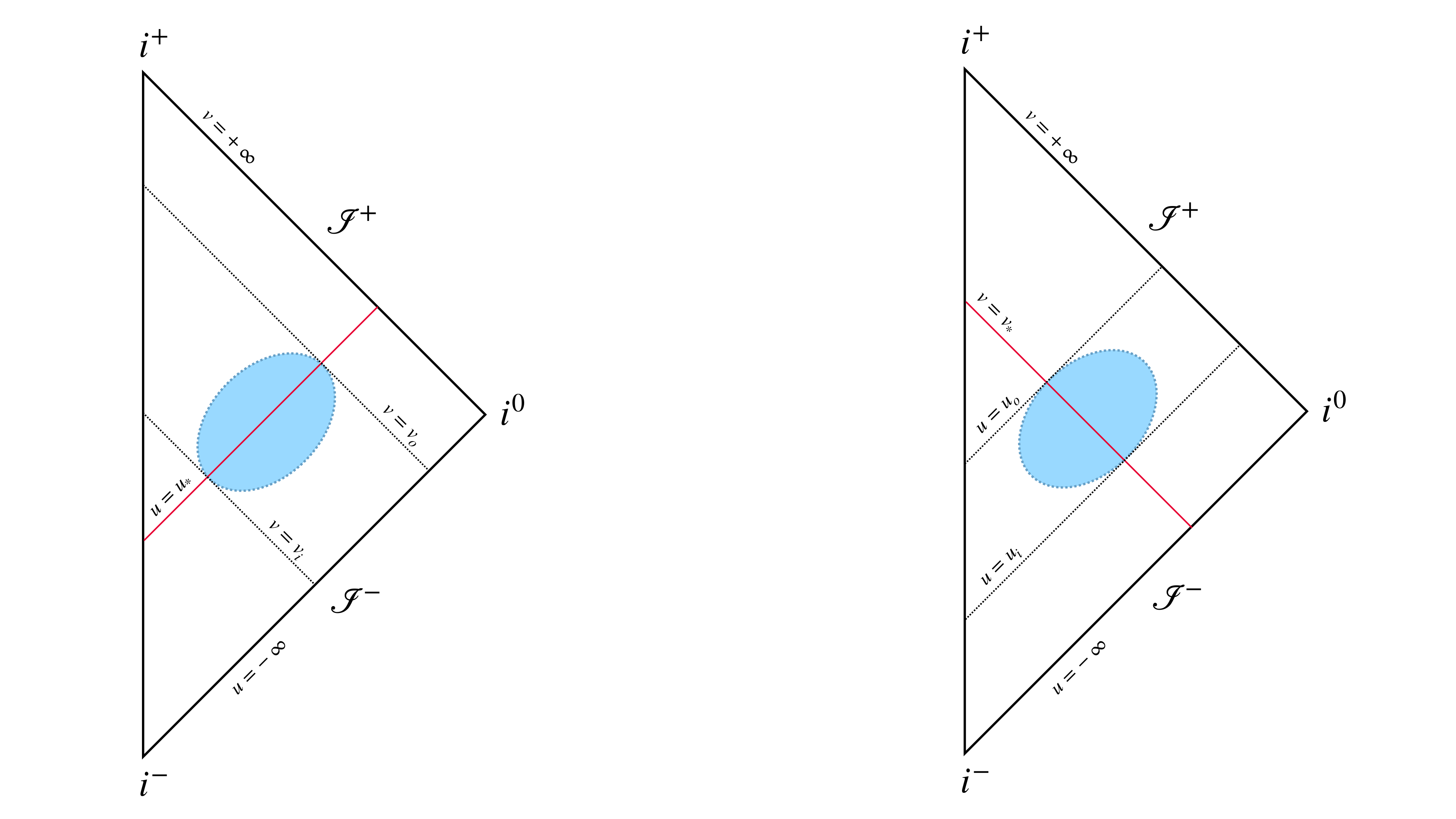}
	\caption{\label{Fig:RBH} Black hole spacetimes. Left plot: A slice of constant $u=u_*$ intersecting the trapping horizon of a black hole spacetime twice, once at $v=v_i$ and once at $v=v_o > v_i$. Right plot: A slice of constant $v=v_*$ intersecting trapping horizon of black hole spacetime  twice, once at $u=u_i$ and once at $u=u_o > u_i$. Note that in these plots the intersections can be arbitrary, i.e., they do not need to mark the maximal extensions of the trapped region in the $u$-direction or $v$-direction, respectively. In the left plot, the NCC is necessarily violated along the red line from the local minimum at $v=v_{\text{min}}$ of $\theta_+(u_*,v)$ which is closest to $v_{o}$, and up to the boundary of the trapped region at $v=v_{o}$. By contrast, in the right plot NCC violation is not necessary along the red line.}
\end{figure}

In order to illustrate the previous statements explicitly, let us now consider a transient black hole and fix a slice of constant $u= u_*$ such that it intersects the boundary of the trapped region twice, once at $v=v_i$ and once at $v=v_o > v_i$, where respectively $\theta_+ = 0$, see the plot on the l.h.s.~in~Fig.~\ref{Fig:RBH}.~\footnote{Notice that in the left plot of Fig.~\ref{Fig:RBH}, $v=v_i$ and $v=v_o$ are also the instants of time at which the geometry is extremal, i.e., the inner and outer horizons coincide as the trapped region opens or closes.}

We are interested in the expansion $\theta_+$ evaluated along this constant-$u$ slice, 
\ba
\theta_+(u_*,v) &=& \frac{2}{R(u_*,v)} \partial_v R(u_*,v)\,,
\ea
and its variation as $v$ increases. 

From the fact that
\ba
\theta_+(u_*,v) &=&
\begin{cases}
	> 0\,, \quad v < v_i\,,\\
	= 0\,, \quad v = v_i\,,\\
	< 0\,, \quad v \in (v_i,v_o)\,,\\
= 0\,, \quad v = v_o\,,\\
> 0\,, \quad v > v_o\,,
\end{cases}
\ea
it follows that $\theta_+$, evaluated along this constant-$u$ slice as a function of $v$, has at least one negative local minimum in the trapped region. Let $v=v_{\text{min}} \in (v_i ,v_{o})$ be the minimum closest to $v_{o}$. Thus, 
\ba
\partial_v \theta_+(u_*,v_\text{min}) &=& 
-\frac{1}{2}\theta_+^{2}(u_*,v_\text{min})- R_{\mu\nu}k_+^\mu k_+^\nu (u_*,v_\text{min}) \,\, = \,\, 0\,,
\ea
and hence at this point the NCC is violated,
\ba
R_{\mu\nu}k_+^\mu k_+^\nu (u_*,v_\text{min}) &=& - \frac{1}{2}\theta_+^{2 }(u_*,v_{\text{min}})\,\, < \,\, 0\,.
\ea
Moreover, the fact that $\theta_+(u_*,v)$ has no further minima in $\qty(v_{\text{min}},v_o)$ implies that this function is monotonically increasing on this interval, such that $\partial_v \theta_+(u_*,v) >0$. Therefore, for any $v \in (v_{\text{min}},v_o)$ it holds,
\ba
R_{\mu\nu}k_+^\mu k_+^\nu (u_*,v) &=& - \frac{1}{2}\theta_+^{2 }(u_*,v) - \partial_v \theta_+ (u_*,v)\,\, < \,\, 0\,,
\ea
and hence the NCC along this constant-$u$ slice is necessarily violated at any $v > v_{\text{min}}$ up to the boundary $v=v_o$ of the trapped region.

We can also compute the total integrated amount of NCC violation within this $v$-interval by using~\footnote{That is, we now consider the averaged null convergence condition (ANCC), the curvature-based analogue of the averaged null energy condition (ANEC) in general relativity. Note that the integral is carried out using an affine parameter.}
\ba
\theta_{+}(u_*,v_o) - \theta_{+}(u_*,v_{\text{min}}) &=& \int_{v_{\text{min}}}^{v_o} \dd{v} \partial_v \theta_+(u_*,v)\nn\\
&=&  - \frac{1}{2}  \int_{v_{\text{min}}}^{v_o} \dd{v} \theta_+^{2}(u_*,v) -  \int_{v_{\text{min}}}^{v_o} \dd{v} R_{\mu\nu}k_+^\mu k_+^\nu (u_*,v)\,,\quad 
\ea
i.e.,
\ba
\int_{v_{\text{min}}}^{v_o} \dd{v} R_{\mu\nu}k_+^\mu k_+^\nu (u_*,v) &=& - \frac{1}{2}  \int_{v_{\text{min}}}^{v_o} \dd{v} \theta_+^{2}(u_*,v) + \theta_+(u_*,v_{\text{min}})\,,
\ea
where we remind the reader that $\theta_+(u_*,v_o)=0$ and that by construction $\theta_+(u_*,v_{\text{min}})<0$.\\

For completeness, let us now instead consider a fixed slice of constant $v=v_*$ such that it intersects the boundary of the trapped region twice, once at $u=u_i$ and once at $u=u_o > u_i$, where respectively $\theta_+ = 0$, see the plot on the r.h.s.~in~Fig.~\ref{Fig:RBH}. Along such a constant-$v$ slice, the NCC does not need to be violated --- even though $\theta_+$ will in general also have at least one negative local minimum at some $u =u_{\text{min}}$ in the trapped region. 
Indeed, without any further information about the past and future evolution in advanced time $v$, i.e., considering the spacetime diagram only along the red line in the plot on the r.h.s.~in Fig.~\ref{Fig:RBH}, is equivalent to just considering an eternal static regular black hole with two horizons. Such a static regular black-hole spacetime may or may not violate the NCC depending, for instance, on whether the regularisation is achieved through an anti-de Sitter or de Sitter core~\cite{Borissova:2025msp,Borissova:2025hmj}.

\subsection{Formation and expansion of an anti-trapped region}\label{SecSub:AntiTrapped}

\begin{figure}[b]
	\centering
	\includegraphics[width=0.95\textwidth]{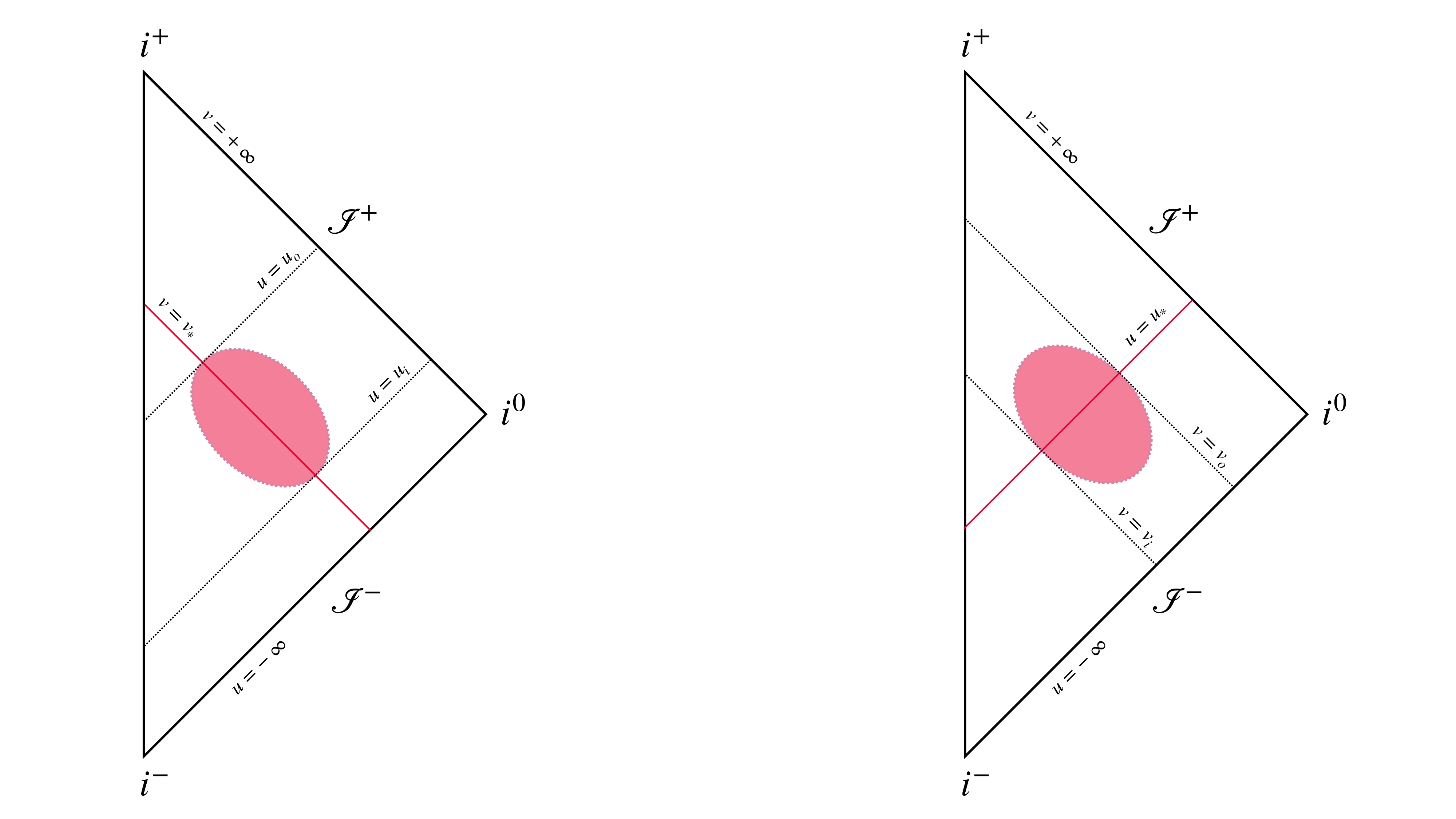}
	\caption{\label{Fig:RWH} White hole spacetimes. Left plot: A slice of constant $v=v_*$ intersecting the anti-trapped region boundary of a white hole spacetime twice, once at $u=u_i$ and once at $u=u_o > u_i$. Right plot: A slice of constant $u=u_*$ intersecting the boundary of the anti-trapped region twice, once at $v=v_i$ and once at $v=v_o > v_i$. Note that in these plots the intersections can be arbitrary, i.e., they do not need to mark the maximal extensions of the anti-trapped region in the $v$-direction or $u$-direction, respectively. In the left plot, the NCC is necessarily violated along the red line from the boundary of the anti-trapped region at $u=u_{i}$ up to the local maximum at $u=u_{\text{max}}$ of $\theta_-(u,v_*)$ which is closest to $u_{i}$. By contrast, in the right plot, NCC violation is not necessary along the red line.}
\end{figure}

We now proceed by illustrating the previous statements concerning the formation of an anti-trapped region explicitly. Thus, let us consider a transient white hole and fix a slice of constant $v=v_*$ such that it intersects the anti-trapped region boundary twice, e.g.~at $u=u_i$ and at $u=u_o >u_i$, where respectively $\theta_- =0$, see the plot on the l.h.s.~in Fig.~\ref{Fig:RWH}.~\footnote{Notice that in the left plot of Fig.~\ref{Fig:RWH}, $u=u_i$ and  $u=u_o$ are also the instants of time at which the geometry is extremal, i.e., the inner and outer horizons coincide as the anti-trapped region opens or closes.}

We are interested in the expansion $\theta_-$ evaluated along this constant-$v$ slice, 
\ba
\theta_-(u,v_*) &=& \frac{2}{R(u,v_*)} \partial_u R(u,v_*)\,,
\ea
and its variation as $u$ increases. 

From the fact that
\ba
\theta_-(u,v_*) &=&
\begin{cases}
	< 0\,, \quad u < u_i\,,\\
	= 0\,, \quad u = u_i\,,\\
	> 0\,, \quad u \in (u_i,u_o)\,,\\
	= 0\,, \quad u = u_o\,,\\
	< 0\,, \quad u > u_o\,,
\end{cases}
\ea
it follows that $\theta_-$ evaluated along this constant-$v$ slice as a function of $u$ has at least one positive local maximum in the anti-trapped region. Let $u=u_{\text{max}} \in (u_i ,u_{o})$ be the maximum closest to $u_{i}$. Thus, 
\ba
\partial_u \theta_-(u_{\text{max}},v_*) &=& 
-\frac{1}{2}\theta_-^{2}(u_{\text{max}},v_*)- R_{\mu\nu}k_-^\mu k_-^\nu (u_{\text{max}},v_*) \,\, = \,\, 0\,,
\ea
and hence at this point the NCC is violated,
\ba
R_{\mu\nu}k_-^\mu k_-^\nu (u_{\text{max}},v_*) &=& - \frac{1}{2}\theta_-^{2 }(u_{\text{max}},v_*)\,\, < \,\, 0\,.
\ea
Moreover, the fact that $\theta_-(u,v_*)$ has no previous maxima in $\qty(u_i,u_{\text{max}})$ implies that this function is monotonically increasing on this interval, such that $\partial_u \theta_-(u,v_*) >0$. Therefore, for any $u \in \qty(u_i,u_{\text{max}})$ it holds,
\ba
R_{\mu\nu}k_-^\mu k_-^\nu (u,v_*) &=& - \frac{1}{2}\theta_-^{2 }(u,v_*) - \partial_u \theta_- (u,v_*)\,\, < \,\, 0\,,
\ea
and hence the NCC along this constant $v$-slice is necessarily violated at any $u < u_{\text{max}}$ from the boundary at $u=u_i$ of the anti-trapped region.
We can also compute the total integrated amount of NCC violation within this $u$-interval by using
\ba
\theta_{-}(u_\text{max},v_*) - \theta_{-}(u_i,v_*)  &=& \int_{u_i}^{u_{\text{max}}} \dd{u}  \partial_u \theta_-(u,v_*)\nn\\
&=&  - \frac{1}{2}  \int_{u_i}^{u_{\text{max}}} \dd{u} \theta_-^{2}(u,v_*) -   \int_{u_i}^{u_{\text{max}}} \dd{u} R_{\mu\nu}k_-^\mu k_-^\nu (u,v_*)\,,\quad 
\ea
i.e.,
\ba
\int_{u_i}^{u_{\text{max}}} \dd{u} R_{\mu\nu}k_-^\mu k_-^\nu (u,v_*) &=& - \frac{1}{2} \int_{u_i}^{u_{\text{max}}} \dd{u} \theta_-^{2}(u,v_*) - \theta_-(u_\text{max},v_*)\,,
\ea
where we remind the reader that $\theta_-(u_i,v_*)=0$ and $\theta_-(u_\text{max},v_*)>0$.\\

By an argument analogous to the one at the end of Sec.~\ref{SecSub:Trapped}, we can also conclude that NCC violation along a fixed slice of constant $u=u_*$ interecting the trapped region at $v=v_i$ and $v=v_o > v_i$, such as in the plot on the r.h.s.~in Fig.~\ref{Fig:RWH}, is not necessary and depends on the specifics of the eternal regular white hole state. 

\section{Kinematic construction of bounces}\label{Sec:KinematicConstruction}

\subsection{Metric in generalised Painlev\'e-Gullstrand coordinates}

For the construction of a kinematical model describing a bounce, we will introduce a simple ansatz for the line element in generalised Painlev\'e-Gullstrand coordinates $(t,r,\theta,\varphi)$ given by
\ba\label{eq:MetricPG}
\dd{s}^2 &=& -\qty[1- n(t,r)^2]\dd{t}^2 + 2 n(t,r)\dd{t}\dd{r} +\dd{r}^2 + r^2\dd{\Omega}^2\,,
\ea
where
\ba\label{eq:n}
n(t,r) &=& \sigma(t) \sqrt{\frac{2 m(r)}{r}}\,.
\ea
A similar ansatz for the parametrisation of dynamical black-hole-to-white hole transitions has been considered e.g.~in~\cite{Hergott:2025elg,Gaur:2023ved,Hergott:2022hjm}.

We assume that the mass function $m(r) > 0$ becomes a constant proportional to the Arnowitt--Deser--Misner mass $M$ of the static counterpart of the spacetime in the limit $r \to \infty$. We will also assume that $m(r)$ vanishes sufficiently fast as $\mathcal{O}\qty(r^3)$ for $r\to 0$, such that the metric does not have a curvature singularity at the core $r=0$. In practice, later, we will simply choose $m(r)$ to be the Misner--Sharp mass function of a de Sitter core regular black hole. The transition function $\sigma(t)$ will be modeled so that $\sigma(t) \to \pm \sigma_0$ for some $\sigma_0 > 0$ representing a local maximum, or local minimum, respectively, at intermediate early and late times within a finite time interval for a bounce. Otherwise we will require $\sigma(t) \to 0$ asymptotically as $t \to -\infty$ and $t\to + \infty$, so that the spacetime becomes Minkowski in this limit. \\

Let us first consider an ansatz for the ingoing and outgoing radial null vectors defined by
\ba
k_-^\mu \,\,=\,\, f_-(t,r)\qty[\qty(\partial_t)^\mu - \qty[n(t,r)+1] \qty(\partial_r)^\mu ]\,, \quad \quad \quad k_+^\mu \,\,= \,\,f_+(t,r) \qty[\qty(\partial_t)^\mu - \qty[n(t,r)-1] \qty(\partial_r)^\mu]\label{eq:kPG}\,,
\ea
where the two scaling functions $f_\pm > 0$ must be determined such that these geodesic vector fields are affine parameterised, i.e., $k_\pm^\nu \nabla_\nu k_\pm^\mu = 0$. This condition results in the two partial differential equations
\ba
\qty(n + 1) \partial_r f_- + f_- \partial_r n - \partial_t f_- &=& 0\,,\\
\qty(n - 1) \partial_r f_+ + f_+ \partial_r n - \partial_t f_+ &=& 0\,.
\ea
An evaluation of the expansions of the null vectors~\eqref{eq:kPG}, after using the previous relations imposed on $f_\pm$, leads to
\ba
\theta_- \,\,=\,\, -\frac{2}{r} f_-(t,r)\qty[n(t,r)+1]\,, \quad \quad \quad \theta_+ \,\, =\,\, - \frac{2}{r} f_+(t,r) \qty[n(t,r) -1]\,.
\ea
Thus, the conditions on the signs of the expansions in a trapped, anti-trapped and untrapped region, respectively, translate into
\ba
\text{trapped:} \quad \theta_- < 0\,, \,\,\, \theta_+  < 0\quad \quad & \to  &\quad \quad n(t,r) \,\,>\,\, 1 \,,\\
\text{anti-trapped:} \quad \theta_- > 0\,, \,\,\, \theta_+  > 0 \quad \quad & \to & \quad \quad n(t,r) \,\,<\,\, -1 \,,\\
\text{untrapped:} \quad \theta_- < 0\,, \,\,\, \theta_+  > 0 \quad \quad  &\to &  \quad \quad n(t,r) \,\, \in \,\, (-1,1)\,.
\ea
In particular, the function $n$ must change sign in order to describe a bounce. Since $m(r)>0$, this can only be achieved by a change of sign of $\sigma(t)$ in the untrapped region as $t$ varies. On the other hand, for a dynamical regular black hole or white hole on its own, $n$ does not need to change sign.
\\

Testing the NCC for the above null vectors requires computation of
\ba
R_{\mu\nu}k^\mu_\pm k^\nu_\pm &=& \frac{2}{r} f_\pm^2(t,r) \partial_t n(t,r) \,\, =  \,\, f_\pm^2(t,r) \; \sqrt{\frac{8m(r)}{r^3}}\; \dot{\sigma}(t) \,\, \,\, \;
\propto  \;\; \dot{\sigma}(t),\, \,.
\ea
Thus, NCC violation in this model is entirely controlled by $\dot\sigma(t)$, the $t$-derivative of $\sigma(t)$. The condition $\sigma(t) \to \pm \sigma_0$ at intermediate early and late times necessary in order to describe a bounce already implies that $\sigma(t)$ must change sign in such a way that $\dot{\sigma}<0$, and thus NCC violation in this case is unavoidable.

\subsection{Bardeen black-hole-to-white-hole bounce}\label{SecSub:BardeenBounce}

\begin{figure}[!h]
	\centering
	\includegraphics[width=0.45\textwidth]{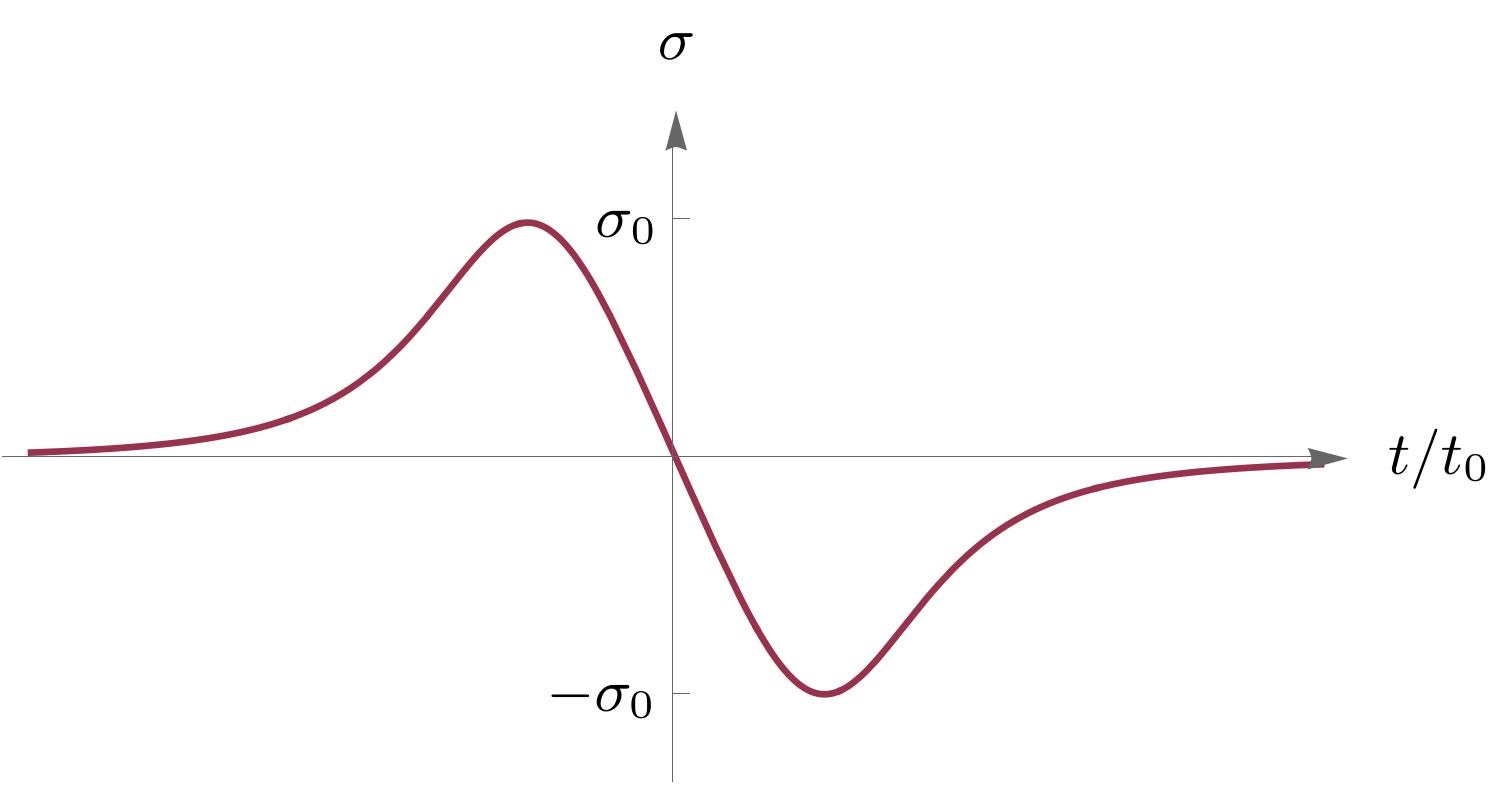}
	\includegraphics[width=0.45\textwidth]{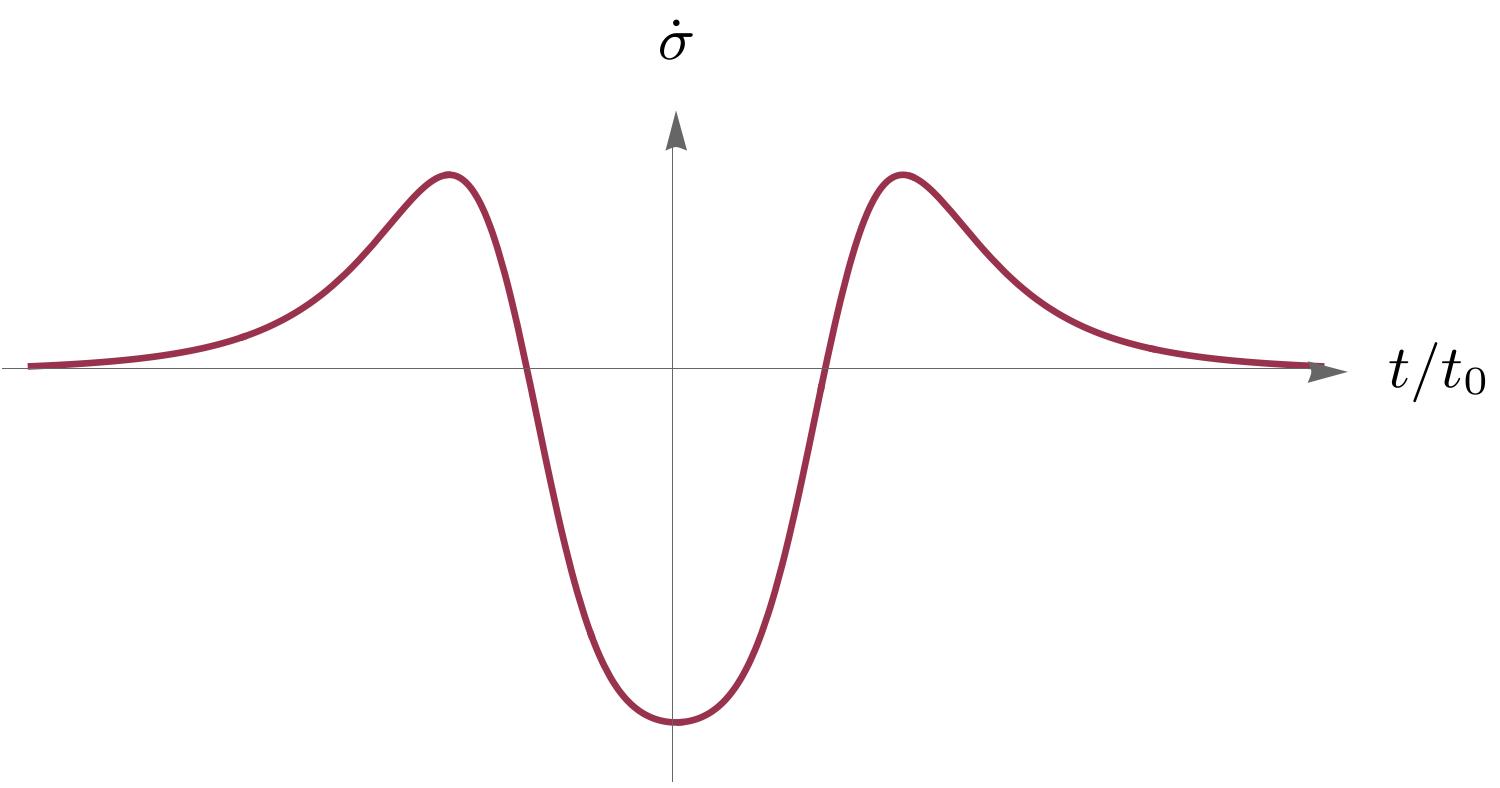}
	\caption{\label{Fig:TransitionFunction} Qualitative behaviour of the transition function $\sigma(t)$, introduced in
Eq.~\eqref{eq:n} and specified in Eq.~\eqref{eq:Sigma}, together with its
derivative. This function is used to model a bounce generated from a regular
black-hole metric. The free parameter has been fixed to $\alpha=1$.
}
\end{figure}

In the following we will consider an exemplary realisation of a bounce starting from a regular black hole metric with a de Sitter core and mass function $m(r) > 0$, and combined with a function $\sigma(t)$ constructed on the interval $t\in (-\infty,\infty)$ such that the model describes a transient black hole followed by a transient white hole. A simple ansatz is
\ba\label{eq:Sigma}
\sigma(t) &=& - \frac{2}{1 + \qty(\frac{t}{t_0})^4} \tanh(\frac{t}{\alpha t_{0}})\,,
\ea
where $t_0$ is a characteristic time scale controlling the width of the transition, and $\alpha >0$ is a dimensionless parameter which
determines the  time scale controlling the speed of the transition. For the main part of this section we will set $\alpha  =1$, but we will comment on the physical implications of varying this parameter later. A qualitative plot of $\sigma$ is shown in Fig.~\ref{Fig:TransitionFunction}.

\begin{figure}[!h]
	\centering
	\includegraphics[width=0.485\textwidth]{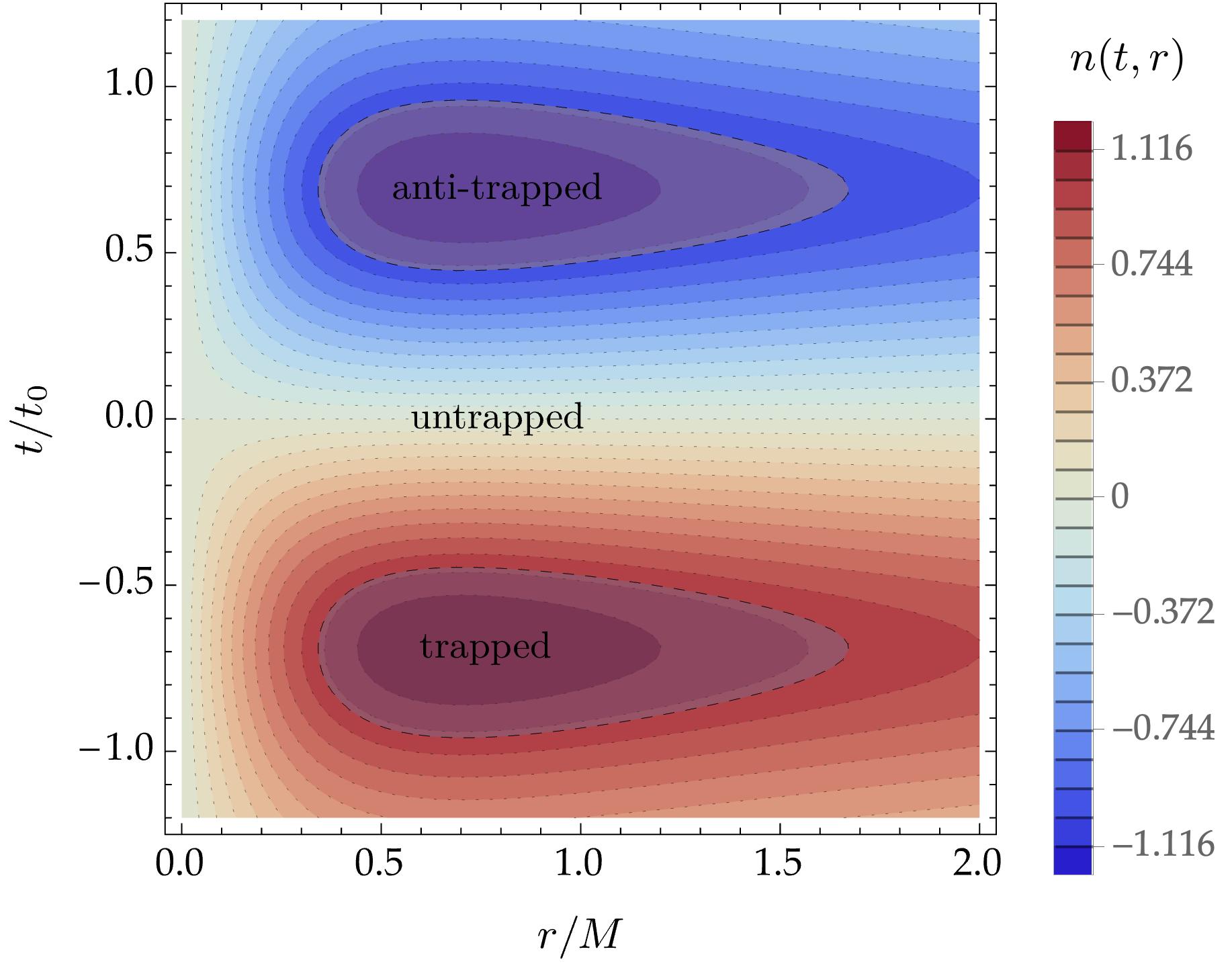}
	\includegraphics[width=0.49\textwidth]{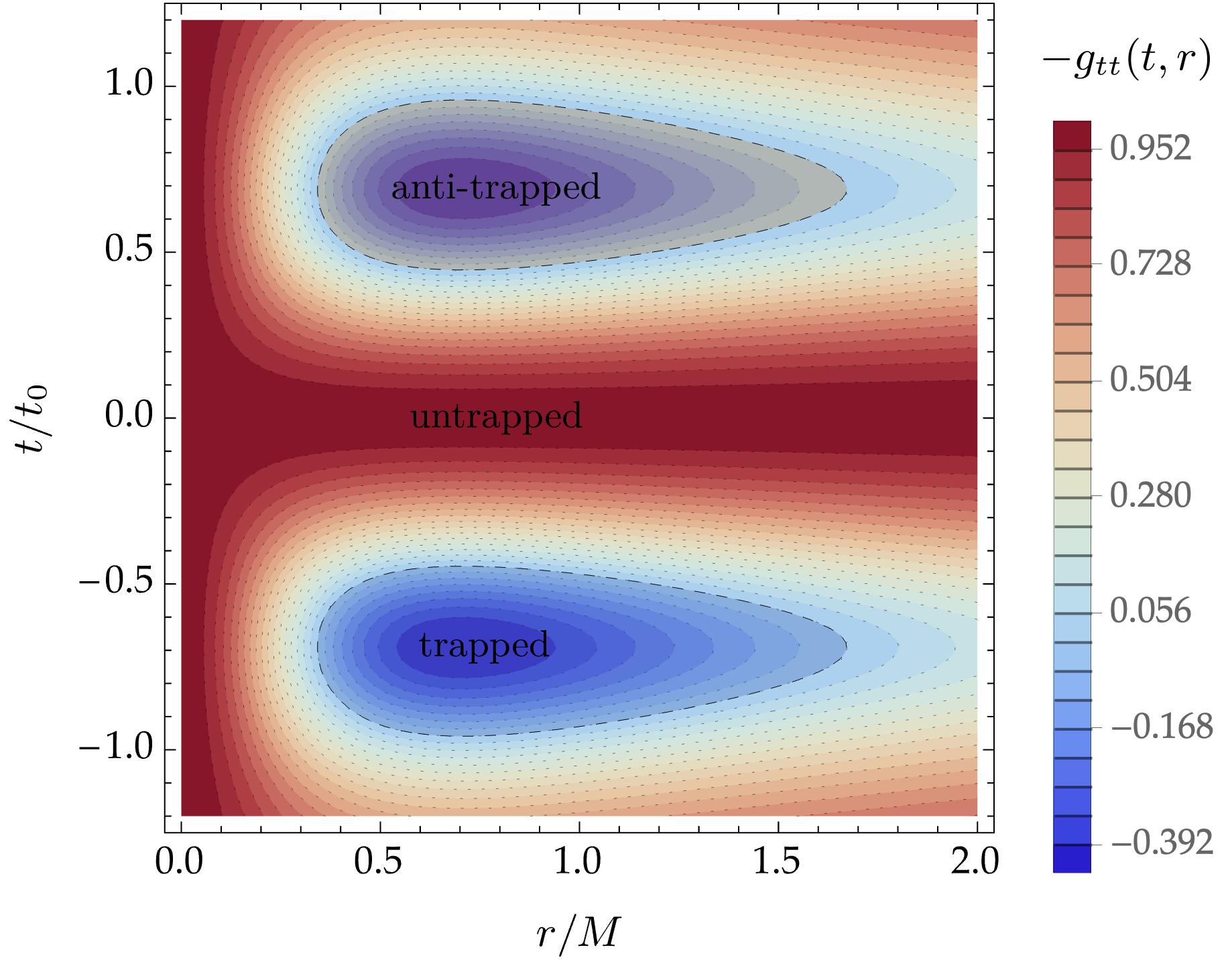}
	\caption{\label{Fig:BardeenBounce} Metric function $n(t,r)$ in~\eqref{eq:n} and $-g_{tt}(t,r) =1-n^2$ for a Bardeen bounce with mass function $m(r)$ in~\eqref{eq:MBardeen} and transition function $\sigma(t)$ in~\eqref{eq:Sigma}, with free parameters set to $\alpha=1$ and $l/M =1/2$. In the trapped region $\theta_\pm <0$, whereas in the anti-trapped region $\theta_\pm >0$. The plot describes a regular black-to-white hole transition.}
\end{figure}

The procedure considered here  of generating a bounce counterpart to a regular black hole can be applied to an arbitrary choice of regular black hole metric with $m(r)>0$.
For concreteness, we will take here $m(r)$ to be the mass function of the Bardeen spacetime~\cite{Bardeen:1968bh},
\ba\label{eq:MBardeen}
m(r) &=& M \qty[\frac{r^3}{\qty[r^2 + l^2]^\frac{3}{2}}]\,,
\ea
where $l$ is a regularisation length parameter chosen to be sufficiently small compared to $M$ such that the static spacetime obtained by setting $\sigma =1$ describes a regular black hole. Fig.~\ref{Fig:BardeenBounce} shows the resulting bounce as indicated by a sign change of the metric function $n(t,r)$, together with the time-time component of the metric. 

\begin{figure}[!h]
	\centering
	\includegraphics[width=0.48\textwidth]{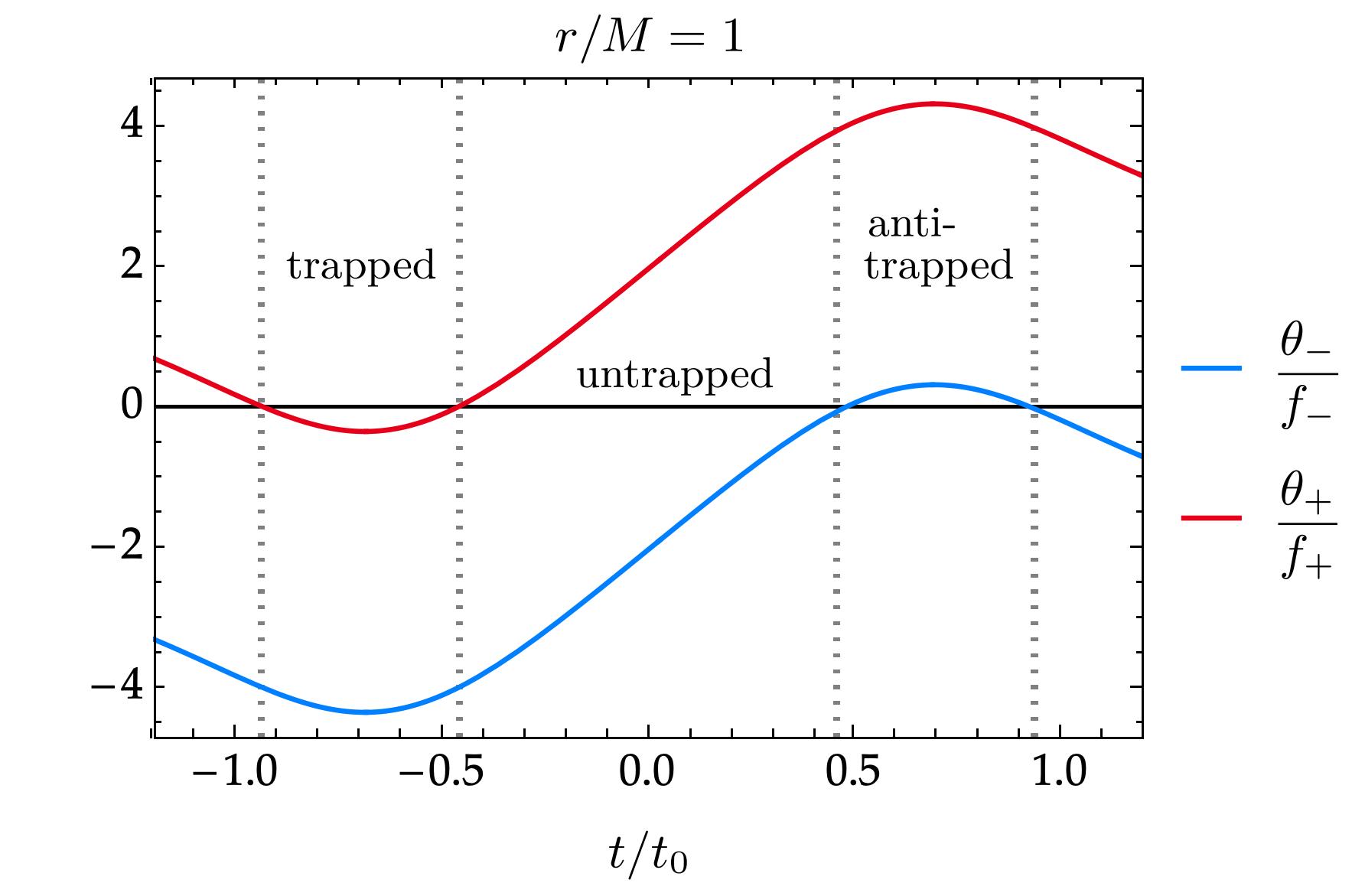}
	\caption{\label{Fig:BardeenBounceTheta} Rescaled expansions~\eqref{eq:ThetaRescaled} as functions of $t/t_0$ along a slice of constant $r/M=1$ for the Bardeen bouncing model with free parameters set to $\alpha=1$ and $l/M =1/2$.}
\end{figure}

Fig.~\ref{Fig:BardeenBounceTheta} shows the rescaled expansions
\ba\label{eq:ThetaRescaled}
\frac{\theta_\pm}{f_\pm} &=& - \frac{2}{r} \qty[n(t,r) \mp 1]
\ea
as functions of $t/t_0$ along a slice of constant $r/M=1$. 

\begin{figure}[!h]
	\centering
	\includegraphics[width=0.49\textwidth]{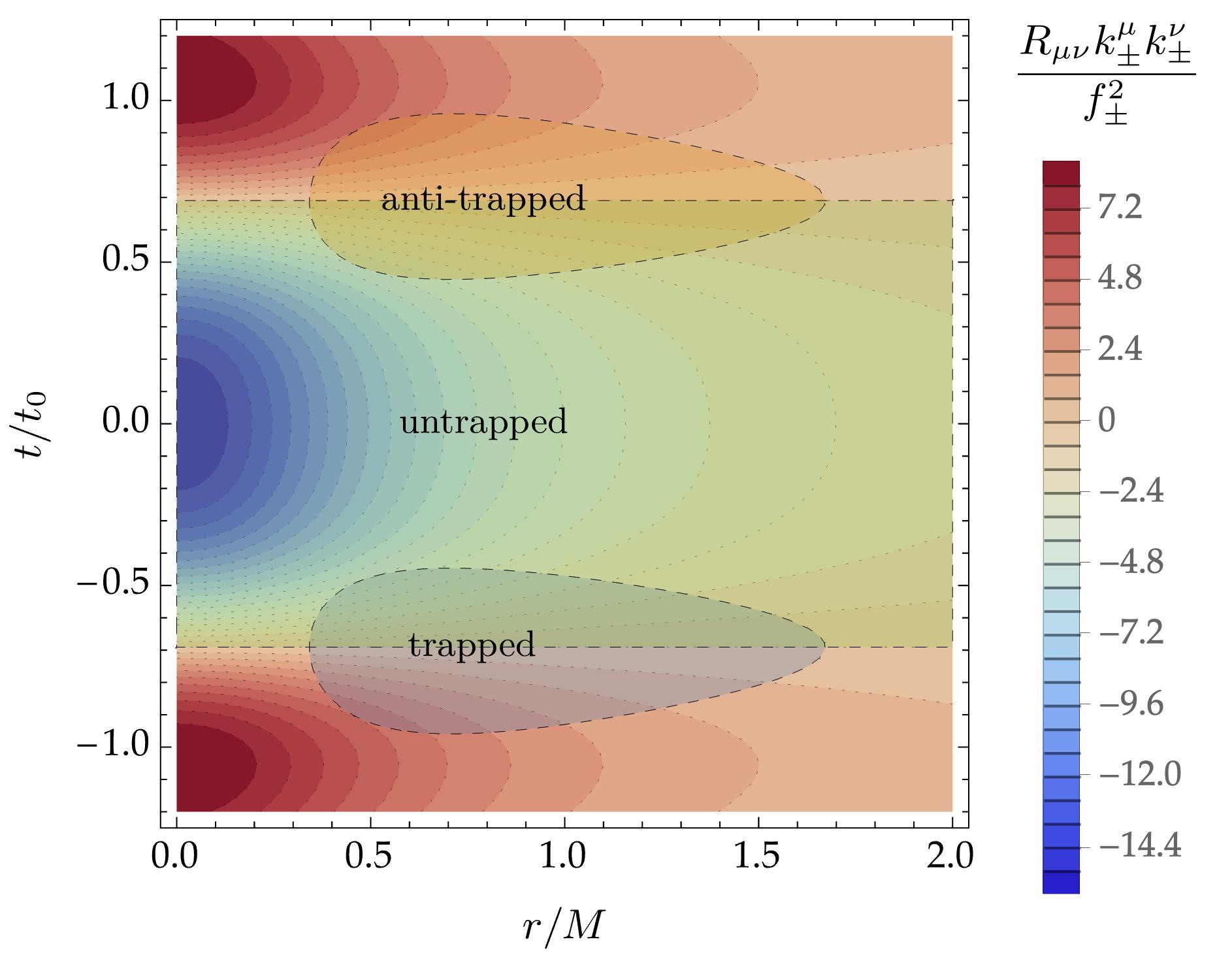}
	\caption{\label{Fig:BardeenBounceNCC} Rescaled NCC function~\eqref{eq:NCCRescaled} for the Bardeen bouncing model with free parameters set to $\alpha=1$ and $l/M =1/2$. The NCC is violated everywhere in the green shaded region, where the trapped region evaporates and the anti-trapped region forms.}
\end{figure}

Finally, Fig.~\ref{Fig:BardeenBounceNCC} shows the rescaled quantity relevant for the NCC,
\ba\label{eq:NCCRescaled}
\frac{R_{\mu\nu}k_\pm^\mu k_\pm^\nu }{f_\pm^2} &=& \frac{2}{r}\partial_t n(t,r) \,\, = \,\, \sqrt{\frac{8m(r)}{r^3}} \dot{\sigma} \,,
\ea
where
\ba\label{eq:SigmaDot}
\dot{\sigma}(t) &=& 
-  \frac{2 t_0^3 }{\qty[t^4+t_0^4]^2 \cosh[2](\frac{t}{\alpha t_0})}\qty[t^4 + t_0^4 - 2 t^3 t_0 \alpha \sinh\qty(\frac{2t}{\alpha t_0})]\,,
\ea
as in the plot on the r.h.s.~in Fig.~\ref{Fig:TransitionFunction}.
We can see that the NCC in this example is violated in a finite intermediate time interval where the trapped region evaporates and eventually disappears, and an anti-trapped region forms and expands, consistently with the conclusions in Sec.~\ref{Sec:PenroseNCC}. The boundaries of NCC violation represent constant-$t$ lines, as we have considered here a simple example in which the transition function $\sigma$ entering the metric through~\eqref{eq:n} is $r$-independent.

\subsection{Hidden wormhole instantaneous-transition limit}\label{SecSub:HiddWorm}

One may also consider the limiting case in which the intermediate untrapped
region separating the black-hole and white-hole phases is shrunk to zero, so
that the trapped and anti-trapped regions touch at a single event. This would
correspond to an instantaneous black-hole-to-white-hole transition. Such a limit is
expected to be associated with a breakdown of the semiclassical continuum
metric description, for the following reason.

Let $t=t_*$ denote the event at which the trapped and anti-trapped regions are
supposed to meet. In our parametrisation, the transition from a trapped to an
anti-trapped region requires the metric function $n(t,r)$ to change from values
larger than unity to values smaller than minus unity,
\ba
\text{trapped:} \quad
\theta_- < 0\,, \qquad \theta_+ < 0
\quad &\Longleftrightarrow& \quad
n(t_*,r)>1\,,
\\
\text{anti-trapped:} \quad
\theta_- > 0\,, \qquad \theta_+ > 0
\quad &\Longleftrightarrow& \quad
n(t_*,r)<-1\,.
\ea
Therefore, if the transition occurs instantaneously, the function $n(t,r)$ must
undergo a discontinuous jump in time. Equivalently, since $n(t,r)$ is
controlled by the transition function $\sigma(t)$, the limiting process
requires $\sigma(t)$ to become infinitely steep at $t=t_*$. In particular,
$\dot{\sigma}(t_*)$ must diverge negatively. It then follows from
Eq.~\eqref{eq:NCCRescaled} that the corresponding violation of the NCC becomes
unbounded.

\begin{figure}[!h]
	\centering
	\includegraphics[width=0.485\textwidth]{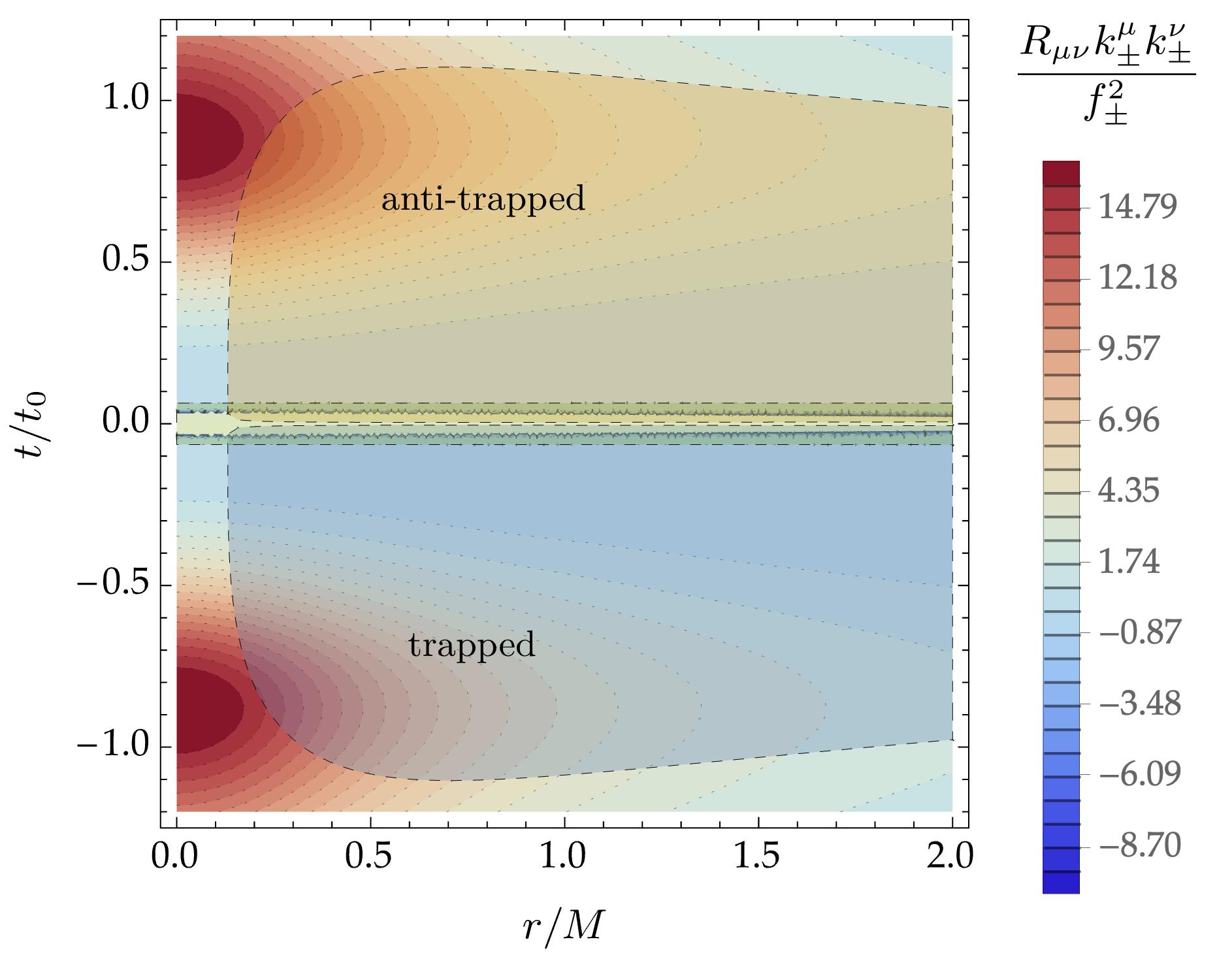}
	\includegraphics[width=0.49\textwidth]{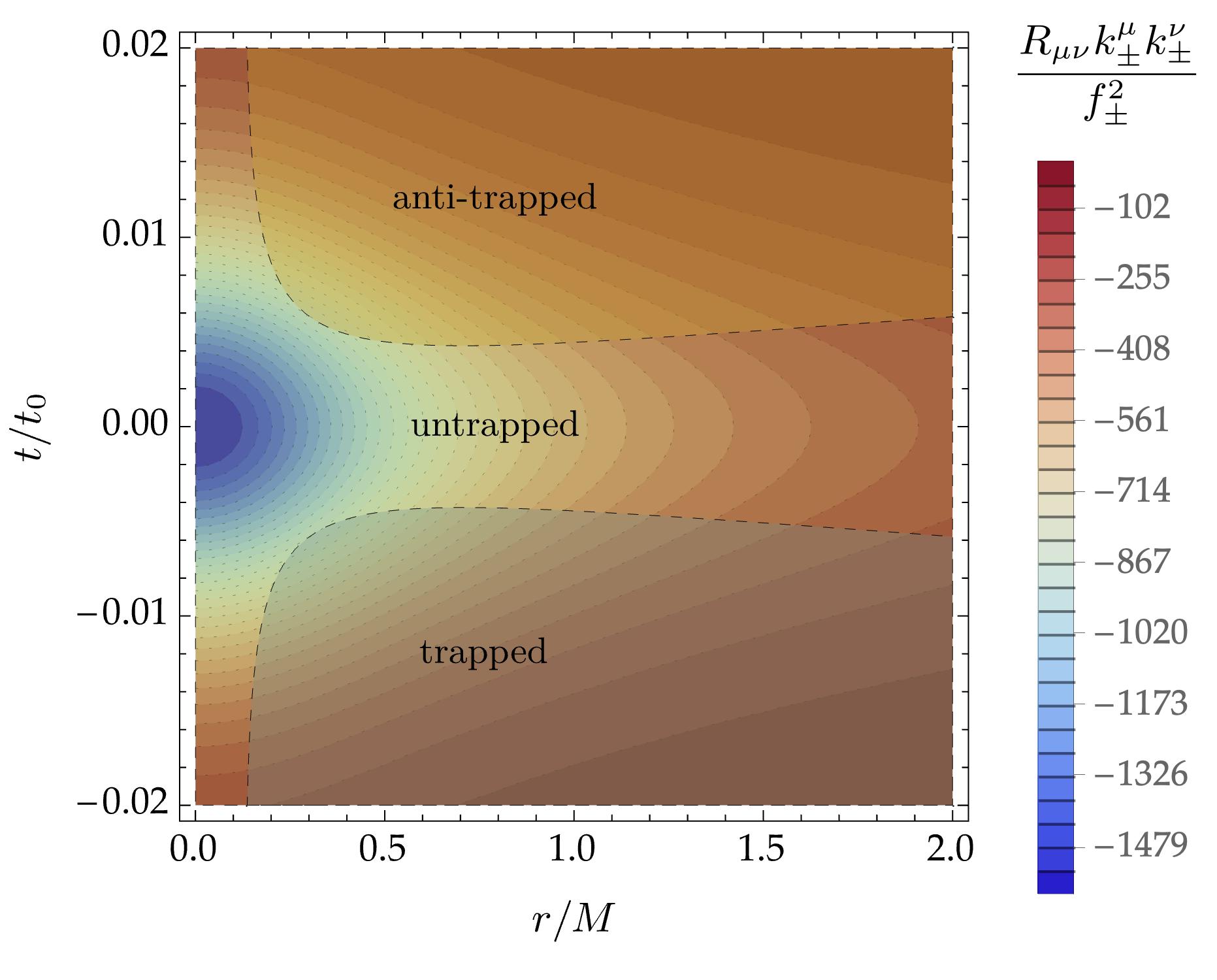}
	\caption{\label{Fig:BardeenBounceNCC2}
	Rescaled NCC function~\eqref{eq:NCCRescaled} for the Bardeen bouncing
	model, with parameters $\alpha=1/100$ and $l/M=1/2$. The NCC is violated
	throughout the green shaded region, corresponding to the dynamical phase in
	which the trapped region evaporates and the anti-trapped region forms. The
	right panel zooms in on the transition region, showing the rapid growth of
	the NCC violation as the instantaneous-transition limit is approached. In
	the limit $\alpha\to0$, the black-hole-to-white-hole transition becomes
	instantaneous; however, this requires the metric to become non-analytic and
	the NCC violation to become unbounded, signalling a potential breakdown of
	the effective semiclassical continuum description.}
\end{figure}

In the explicit model considered in this section, the instantaneous-transition
limit is obtained by taking $\alpha\to0$. For $t_*=0$, the transition function
then becomes infinitely steep. Indeed, Eq.~\eqref{eq:SigmaDot} gives
\ba
\lim_{\alpha\to0}\dot{\sigma}(0)
=
-\lim_{\alpha\to0}\frac{2}{\alpha t_0}
=
-\infty\,.
\ea
Consequently, the quantity controlling the NCC in
Eq.~\eqref{eq:NCCRescaled} becomes unbounded along the constant-time slice
$t=0$, at any finite value of $r/M$. This behaviour signals the breakdown of
the effective semiclassical description and suggests that an instantaneous
black-hole-to-white-hole transition would require a fully quantum-gravitational
treatment, for instance along the lines discussed in
Ref.~\cite{Soltani:2021zmv}. Fig.~\ref{Fig:BardeenBounceNCC2} illustrates
this behaviour for $\alpha=1/100$.
\newpage
\section{Discussion}\label{Sec:Discussion}

There are only a few candidate geometries that can replace a dynamical
singular black or white hole by a dynamical, non-singular spacetime containing
a trapping or anti-trapping horizon~\cite{Carballo-Rubio:2019nel,Carballo-Rubio:2019fnb}.
In this work we have focused on bouncing geometries, in which a dynamical
regular black hole evolves into a dynamical regular white hole. We have shown
that such a transition necessarily entails a violation of the null convergence
condition (NCC), as follows directly from the logic of the Penrose theorem.
While the NCC is equivalent to the null energy condition (NEC) in general
relativity, it is a purely geometric condition and does not rely on any
particular choice of gravitational field equations. Our conclusions are
therefore theory agnostic: they depend only on the kinematics of the spacetime
geometry, and not on the microscopic mechanism responsible for singularity
resolution.~\footnote{Indeed, singularity resolution in black-hole and
cosmological spacetimes can be achieved through purely gravitational
mechanisms, see, e.g.,~\cite{Bueno:2024dgm,Borissova:2026wmn,  Bueno:2025qjk,Borissova:2026krh,Frolov:2024hhe,Colleaux:2017ibe,Colleaux:2019ckh,DiFilippo:2024mwm,Tsuda:2026xjc,Colleaux:2026qew,Liu:2026xqf,Frolov:2026rcm,Hennigar:2025yqm,Aguayo:2025xfi,Borissova:2026rbi,Bakhoda:2026olq} for regular black-hole vacuum solutions, and, e.g.,~\cite{Bueno:2024zsx,Bueno:2024eig,Bueno:2025gjg,Bueno:2025zaj,Bueno:2026dln,PinedoSoto:2026hfm,
Borissova:2026klg,Sueto:2026epz,Borissova:2026prd,Borissova:2026kwu} for the coupling of such theories to matter resulting in non-singular black holes and cosmologies without a violation of energy conditions.}

The NCC violation found in bouncing geometries has a simple kinematical
interpretation. It reflects the defocusing of null geodesics required in two
distinct stages of the evolution: the closing of the trapped region and the
opening of the anti-trapped region. The formation of a black hole does not, by
itself, require a violation of the NCC. By contrast, the evaporation and
eventual disappearance of a trapped region does, independently of the specific, quantum or otherwise, mechanism responsible for the evaporation. The same
statement applies, by time reversal, to the formation of a white hole: the
appearance of an anti-trapped region also requires null geodesic defocusing.
In this respect, the NCC violation in a bounce is closely related to that
encountered in a one-way hidden wormhole, where a trapped region is joined
directly to an anti-trapped region. The crucial difference is that, in the
wormhole case, the transition is instantaneous and the metric is expected to
become non-analytic at isolated points. Our analysis supports the expectation
that the limiting case of an instantaneous black-hole-to-white-hole transition is
associated with a breakdown of the effective continuum metric description, and
therefore requires genuinely quantum-gravitational input.

It is important to distinguish this dynamical violation of the NCC from the
violation of the timelike convergence condition (TCC) already present in
static regular black-hole geometries. The NCC violation discussed here is tied
to the kinematics of the transition itself: the evaporation of the trapped
region and the formation of the anti-trapped region. It is not directly the
same effect as the TCC violation required to avoid the focusing point in the
Hawking--Penrose theorem~\cite{Hawking:1970zqf,Hawking:1973uf};
see, e.g., Ref.~\cite{Borissova:2025hmj}. Thus, even when a static regular
black hole already evades singularity formation through a violation of the
TCC, a further and distinct NCC violation is required once the spacetime is
made dynamical in such a way as to produce a black-hole-to-white-hole bounce.

To illustrate these general statements, we have developed a systematic
procedure for constructing bouncing geometries starting from static regular
black holes with a de Sitter core. The Bardeen spacetime was used as a
representative example, but the construction is not tied to this particular
mass function. Static regular black holes with de Sitter cores need not
violate the NCC~\cite{Borissova:2025msp,Borissova:2025hmj}. Nevertheless,
their dynamical bouncing counterparts inevitably do. This conclusion is
independent of the detailed choice of time-dependent interpolation function:
the sign change needed to turn a trapped region into an anti-trapped region
forces a phase in which the NCC is violated.

If such bouncing geometries are interpreted as solutions of the Einstein
equations, the required NCC violation translates directly into a violation of
the NEC by the matter source (see, e.g.,~\cite{Wang:2026jvo,Wang:2026sqr} for an analysis of energy conditions for spherically symmetric spacetimes).
In modified theories of gravity
the question of energy conditions versus convergence is however ambiguous
and this reflects a broader gap in our understanding of 
singularity theorems beyond
general relativity. One of the main challenges in addressing this question are higher derivatives or non-minimal couplings which prevent a straightforward assigment of an effective matter stress-energy tensor to which energy conditions could be applied. In fact, from the point of view of the singularity theorems, energy conditions in modified gravity theories would be meaningful only if they are defined to apply to the effective energy-momentum tensor identified so that $G_{\mu\nu} = 8 \pi T_{\mu\nu}^{\text{eff}}$, which itself may contain separate gravitational and matter terms or combined gravity-matter contributions originating from non-minimal couplings. Only with this definition would energy conditions be equivalent to the convergence conditions, the latter being ultimately the only relevant and purely geometric conditions in applications of  the theorems.

There are, however, reasons to expect that purely matter-sector energy-condition
violations may still play an important role in dynamical bouncing solutions of
modified gravity theories that admit static regular black-hole or white-hole
vacua.  Generic classes of such static metrics can be realised as solutions of
effective field equations constructed from two-dimensional Horndeski
theory~\cite{Carballo-Rubio:2025ntd}. Moreover, these two-dimensional
Horndeski theories can themselves be obtained from the spherical reduction of
purely gravitational theories in $d=4$ or higher
dimensions~\cite{Borissova:2026krh,Borissova:2026wmn,Colleaux:2019ckh,Colleaux:2017ibe}.
Thus, realising a given static regular black-hole or white-hole metric as a
solution of the effective second-order field equations on the space of
spherically symmetric metrics constructed in
Ref.~\cite{Carballo-Rubio:2025ntd} amounts to realising it as a vacuum
solution of an appropriate higher-dimensional gravitational
theory~\cite{Borissova:2026krh,Borissova:2026wmn}. 

The gravitational theories whose spherical reductions yield general integrable two-dimensional Horndeski theories are known as general quasi-topological gravities~\cite{Borissova:2026krh,Borissova:2026wmn,Borissova:2026rbi}. Not only is the spherically symmetric sector of these theories second order, but moreover vacuum solutions are static and determined by a single function $g_{tt} = -f(r)$ in Schwarzschild gauge which satisfies an algebraic equation~\cite{Borissova:2026krh,Borissova:2026wmn}. This makes it possible to evaluate convergence conditions exactly on shell~\cite{Borissova:2026prd}.

In particular, for the spherical sector of these theories, there exists a direct translation between violations of convergence conditions reformulated dynamically onshell for regular black-hole spacetimes sourced as quasi-topological vacuum solutions  on the one hand, and the required violation of energy conditions when these same spacetimes are sourced as effective generalised static Vaidya solutions of general relativity on the other~\cite{Borissova:2026prd}. In other words, the regularisation of static geometries in these modified quasi-topological gravities is carried by the gravitational sector, while a time dependence allowing for the description of geometries  of the type considered here can be achieved by a minimal coupling, so that energy conditions for dynamical geometries can be addressed.

Thus, once the regularisation has been geometrically implemented through the
modified gravitational sector, one can introduce dynamics by coupling the
system, for example, to Vaidya-type sources~\cite{Boyanov:2025pes}; see
Ref.~\cite{Borissova:2026dlz,Borissova:2026prd} for an application. 
Such sources would provide the time-dependent transition function required for
a bounce, corresponding in our construction to the function $\sigma(t)$.
Since this same function controls the unavoidable NCC violation identified in
the present work, and since it would be sourced by matter in such a dynamical
realisation, it is natural to expect that matter violating the NEC will also
be required in dynamical black-hole-to-white-hole bounces within modified gravity,
even when the associated static regular black-hole and white-hole geometries
exist as vacuum solutions of the modified theory. We leave this issue to future investigations.

\begin{acknowledgments}
	
The work of JB is supported by STFC Consolidated Grant ST/X000575/1.

\end{acknowledgments}

\appendix

\section*{Appendix}\label{Appendix}

One may object that~\eqref{eq:MetricDoubleNull} is not the most general spherically symmetric line element and accordingly the discussion in Sec.~\ref{Sec:BounceDefinition} would have to be extended to describe general spherically symmetric spacetimes in double-null coordinates $\qty(u,v,\theta,\varphi)$ given by
\ba\label{eq:Metricf}
\dd{s}^2 &=& - f(u,v)\dd{u}\dd{v} + r(u,v)^2 \dd{\Omega}^2\,.
\ea
It is well-known that the conformal transformation bringing~\eqref{eq:Metricf} into the form~\eqref{eq:MetricDoubleNull} is singular at spacetime points where $f=0$, and this is indeed an issue if $u$ and $v$ are the standard advanced and retarded double-null coordinates. For instance, the Schwarzschild metric in these coordinates is given by
\ba
\dd{s}^2 &=& - \qty(1-\frac{2m}{r})\dd{u}\dd{v} + r(u,v)^2 \dd{\Omega}^2\,,
\ea
where $r(u,v)$ is defined implicitly in terms of $u$ and $v$ by transforming the Schwarzschild radial coordinate $r$ to the tortoise coordinate $r_*$ defined by $\dd{r_*} = f(r)^{-1}\dd{r}$ and identifying $u = t - r_*$ and $v = t + r_*$. The  coordinate singularity at $r = 2m$ corresponding to $f=0$ of the above metric can be avoided by transforming to Kruskal-Szekeres double-null coordinates $U $ and $V$, in terms of which the metric becomes
\ba\label{eq:MetricfKruskal}
\dd{s}^2 &=& - f(U,V)\dd{U}\dd{V} + r(U,V)^2 \dd{\Omega}^2
\ea
with a function $f$ which is manifestly non-zero. One may now apply a conformal transformation to this metric and obtain a line element of the form~\eqref{eq:MetricDoubleNull} which is general enough to analyse the behavior of null geodesics of the original metric~\eqref{eq:MetricfKruskal}. In this sense we consider the metric~\eqref{eq:MetricDoubleNull} to be the one arising through a conformal transformation from a Kruskalised metric of the form~\eqref{eq:MetricfKruskal}. That global Kruskal coordinates are not restricted to single-horizon spacetimes but can be defined also for double-horizon spacetimes has been discussed in~\cite{Soltani:2023tfr}, and we implicitly assume the spacetime metric to be given in such a coordinate system as a starting point of our discussion in Sec.~\ref{Sec:BounceDefinition}.

\enlargethispage{20pt}
\bibliographystyle{jhep}
\bibliography{references}

\end{document}